\documentclass[preprintnumbers,twocolumn,unsortedaddress,superscriptaddress,amsmath,amssymb,prb,nobibnotes,altaffilletter]{revtex4}

\usepackage{graphics,graphicx}
\usepackage{dcolumn}
\preprint{Physics of Fluids}

\begin{document}

\title{Large-scale lognormal fluctuations in turbulence velocity fields}

\author{Hideaki Mouri}
\email{hmouri@mri-jma.go.jp}
\affiliation{Meteorological Research Institute, Nagamine, Tsukuba 305-0052, Japan}

\author{Akihiro Hori}
\altaffiliation[Also at ]{Meteorological and Environmental Sensing Technology, Inc., Nanpeidai, Ami 300-0312, Japan}
\affiliation{Meteorological Research Institute, Nagamine, Tsukuba 305-0052, Japan}

\author{Masanori Takaoka}
\affiliation{Department of Mechanical Engineering, Doshisha University, Kyotanabe, Kyoto 610-0321, Japan}

\begin{abstract}
For several flows of laboratory turbulence, we obtain long records of velocity data. These records are divided into numerous segments. In each segment, we calculate the mean rate of energy dissipation, the mean energy at each scale, and the mean total energy. Their values fluctuate significantly among the segments. The fluctuations are lognormal, if the segment length lies within the range of large scales where the velocity correlations are weak but not yet absent. Since the lognormality is observed regardless of the Reynolds number and the configuration for turbulence production, it is expected to be universal. The likely origin is some multiplicative stochastic process related to interactions among scales through the energy transfer.

\end{abstract}
\maketitle

\section{INTRODUCTION} \label{s1}

Turbulence spans a wide range of scales, from small scales in the dissipation and inertial subranges to large scales in the energy-containing subrange. While the small scales have been studied in detail,\cite{sa97} the large scales have not thus far. There should remain unknown features that are crucial to understand the basic physics of turbulence. Here, we reveal that the rate of energy dissipation, the energy at each scale, and the total energy fluctuate significantly over large scales in lognormal distributions. This lognormality is expected to be universal because it is observed regardless of the Reynolds number and the configuration for turbulence production.

Landau\cite{ll59} was the first to discuss such a large-scale fluctuation. The discussion was that the local rate of energy dissipation $\varepsilon$ should fluctuate significantly in space over large scales.

Obukhov\cite{o62} also discussed the large-scale $\varepsilon$ fluctuation. The $\varepsilon$ fluctuation was assumed to be lognormal. This is because the fluctuation of any positive quantity could be approximated by a lognormal distribution with appropriate values for the average and standard deviation.

The large-scale fluctuations are known to exist on some level,\cite{po97,c03} but their details are not known. Experimentally or numerically, any detailed study requires long data that are composed of numerous ensembles. Each ensemble is required to cover smallest to largest scales of turbulence. Such long data have not been available. The situation is nevertheless improving for laboratory experiments, owing to improvements of the measurement technologies.

We have accordingly undertaken experimental studies of large-scale fluctuations.\cite{m06,m08} By using long velocity data obtained in a wind tunnel, we have shown that the large-scale fluctuations are always significant.\cite{m06} Their standard deviations are comparable to the averages. Also for a case of grid turbulence, we have shown that the large-scale fluctuations of the energy dissipation rate, the energy at each scale, and the total energy are lognormal.\cite{m08}

The next problem is whether the large-scale lognormal fluctuations exist in other turbulent flows. For grid turbulence, boundary layers, and jets at different Reynolds numbers, we show that the lognormal fluctuations do exist. The standard deviations are almost the same among the flows. It is thereby implied that the lognormality is not a mere approximation\cite{o62} but is due to some common process. We discuss this and other implications.

\begingroup
\squeezetable
\begin{table*}
\caption{\label{t1} 
Experimental conditions and turbulence parameters for grid turbulence (G1 and G2), boundary layers (B1 and B2), and jets (J1 and J2). The velocity derivative was obtained as $\partial _x v = [ 8v(x+\delta x)-8v(x-\delta x)-v(x+2\delta x)+v(x-2\delta x) ] / 12\delta x$ with $\delta x=U/f_s$.}

\begin{ruledtabular}
\begin{tabular}{lllcccccc}
\noalign{\smallskip}
Quantity                     &                                         & Units            & G1     & G2     & B1     & B2     & J1      & J2    \\ 
\hline
\noalign{\smallskip}
Measurement position         & $x$                                     & m                &$+1.5$  &$+2.0$  &$+12.5$ &$+12.5$ &$+15.5$  &$+15.5$\\
Measurement position         & $z$                                     & m                &$1.00$  &$1.00$  &$0.35$  &$0.25$  &$0.40$   &$0.40$ \\
Mean streamwise velocity     & $U$                                     & m\,s$^{-1}$      &$12.75$ &$21.16$ &$3.118$ &$11.34$ &$11.48$  &$17.38$\\
Sampling frequency           & $f_s$                                   & kHz              &$40$    &$70$    &$10$    &$60$    &$44$     &$70$   \\
Kinematic viscosity          & $\nu$                                   & cm$^2$\,s$^{-1}$ &$0.143$ &$0.142$ &$0.138$ &$0.143$ &$0.139$  &$0.139$\\
Mean energy dissipation rate & $\langle \varepsilon \rangle= 15 \nu \langle (\partial _x v)^2 \rangle /2$ 
                                                                       & m$^2$\,s$^{-3}$  &$2.81$  &$7.98$  &$0.244$ &$12.6$  &$2.60$   &$7.52$ \\ 
Kolmogorov velocity          & $u_K = (\nu \langle \varepsilon \rangle)^{1/4}$
                                                                       & m\,s$^{-1}$      &$0.0796$&$0.103$ &$0.0428$&$0.116$ &$0.0776$ &$0.101$\\
Rms velocity fluctuation     & $\langle u^2 \rangle^{1/2}$             & m\,s$^{-1}$      &$0.696$ &$1.10$  &$0.552$ &$2.37$  &$1.56$   &$2.34$ \\
Rms velocity fluctuation     & $\langle v^2 \rangle^{1/2}$             & m\,s$^{-1}$      &$0.683$ &$1.06$  &$0.464$ &$1.96$  &$1.36$   &$2.06$ \\
Skewness factor              & $\langle u^3 \rangle / \langle u^2 \rangle^{3/2}$         &&$+0.08$ &$+0.06$ &$-0.22$ &$-0.10$ &$-0.04$  &$-0.04$\\
Skewness factor              & $\langle v^3 \rangle / \langle v^2 \rangle^{3/2}$         &&$-0.01$ &$-0.01$ &$+0.01$ &$-0.01$ &$+0.01$  &$+0.01$\\
Flatness factor              & $\langle u^4 \rangle / \langle u^2 \rangle^2$             &&$3.00$  &$3.02$  &$2.69$  &$2.69$  &$2.60$   &$2.59$ \\
Flatness factor              & $\langle v^4 \rangle / \langle v^2 \rangle^2$             &&$2.98$  &$3.00$  &$3.06$  &$3.05$  &$3.05$   &$3.06$ \\
Kolmogorov length            & $\eta = (\nu ^3 / \langle \varepsilon \rangle )^{1/4}$ 
                                                                       & cm               &$0.0180$&$0.0138$&$0.0322$&$0.0123$&$0.0179$&$0.0137$\\
Taylor microscale            & $\lambda = [2 \langle v^2 \rangle / \langle (\partial _x v)^2 \rangle ]^{1/2}$
                                                                       & cm               &$0.597$ &$0.548$ &$1.35$  &$0.806$ &$1.21$  &$1.08$  \\
Correlation length           & $L_u = \int^{\infty}_{0} \langle u(x+r)u(x) \rangle dr / \langle u^2 \rangle$
                                                                       & cm               &$17.5$  &$17.9$  &$49.0$  &$43.0$  &$128.$   &$124.$ \\
Correlation length           & $L_v = \int ^{\infty}_{0} \langle v(x+r)v(x) \rangle dr / \langle v^2 \rangle$
                                                                       & cm               &$4.46$  &$4.69$  &$6.94$  &$5.68$  &$10.2$   &$10.2$ \\
Microscale Reynolds number   & Re$_{\lambda} = \langle v^2 \rangle^{1/2} \lambda / \nu$  &&$285$   &$409$   &$454$   &$1103$  &$1183$  &$1603$  \\

\end{tabular}
\end{ruledtabular}
\end{table*}
\endgroup

\section{EXPERIMENTS} \label{s2}

Table \ref{t1} shows experimental conditions and turbulence parameters for grid turbulence (G1 and G2), boundary layers (B1 and B2), and jets (J1 and J2). The dataset G2 is from our previous work.\cite{m08} We note that G2 and B1 as well as B2 and J1 have similar values of the microscale Reynolds number Re$_{\lambda}$.

\subsection{Wind tunnel and anemometer} \label{s2a}

The experiments were carried out in a wind tunnel of the Meteorological Research Institute. We adopt coordinates $x$, $y$, and $z$ in the streamwise, spanwise, and floor-normal directions. The corresponding flow velocities are $U+u$, $v$, and $w$. Here $U$ is the average while $u$, $v$, and $w$ are the fluctuations. The origin $x = y = z = 0$\,m is on the floor center at the upstream end of the test section of the wind tunnel. Its size was $\delta x = 18$\,m, $\delta y = 3$\,m, and $\delta z = 2$\,m. These values of $\delta y$ and $\delta z$ remained the same also to $x = -4$\,m.

The wind tunnel had an air conditioner. If necessary, the conditioner was used to constrain the variation of air temperature. The resultant variation was $\pm 1\,^{\circ}$C at most in each experiment, where the kinematic viscosity $\nu$ is assumed to have been constant.

To simultaneously measure $U+u$ and $v$, we used a hot-wire anemometer. The anemometer was composed of a constant temperature system and a crossed-wire probe. The wires were made of platinum-plated tungsten, 5\,$\mu$m in diameter, 1.25\,mm in sensing length, 1\,mm in separation, oriented at $\pm 45^{\circ}$ to the streamwise direction, and 280\,$^{\circ}$C in temperature.

\subsection{Flow configurations} \label{s2b}

For the grid turbulence, we placed a grid at $x = -2$\,m across the flow passage to the test section of the wind tunnel. The grid was composed of two layers of uniformly spaced rods, with axes in the two layers at right angles. The cross section of the rod was $0.04 \times 0.04$\,m$^2$. The spacing of the axes of adjacent rods was 0.20\,m. We set the incoming flow velocity to be 12\,m\,s$^{-1}$ (G1) or 20\,m\,s$^{-1}$ (G2). The measurement position was on the tunnel axis, $y = 0$\,m and $z = 1.00$\,m.

For the boundary layers, roughness blocks were placed over the entire floor of the test section. The block size was $\delta x = 0.06$\,m, $\delta y = 0.21$\,m, and $\delta z = 0.11$\,m. The spacing of the centers of adjacent blocks was $\delta x = \delta y = 0.50$\,m. We set the incoming flow velocity to be 4\,m\,s$^{-1}$ (B1) or 16\,m\,s$^{-1}$ (B2). The measurement position was in the log-law sublayer at $x = +12.5$\,m and $ y = 0$\,m, where the boundary layer had the displacement thickness of 0.2\,m and the 99\% velocity thickness of 0.8\,m.

For the jets, we placed a contraction nozzle. Its exit was at $x = -2$\,m and was rectangular with the size $\delta y = 2.1$\,m and $\delta z = 1.4$\,m. The center was on the tunnel axis. We set the flow velocity at the nozzle exit to be 16\,m\,s$^{-1}$ (J1) or 24\,m\,s$^{-1}$ (J2). The measurement position was at $x = +15.5$\,m, $y = 0$\,m, and $z = 0.40$\,m.\cite{note1}

The skewness factor $\langle v^3 \rangle / \langle v^2 \rangle^{3/2}$ and the flatness factor $\langle v^4 \rangle / \langle v^2 \rangle^2$ at each of the measurement positions were close to the Gaussian values of 0 and 3. Here $\langle \cdot \rangle$ denotes an average. These Gaussian values imply that turbulence had been fully developed and various eddies filled the space randomly and independently.\cite{m02,m03} Not always close to the Gaussian values were $\langle u^3 \rangle / \langle u^2 \rangle^{3/2}$ and $\langle u^4 \rangle / \langle u^2 \rangle^2$. They tend to be sensitive to largest scale motions induced by either the grid, roughness, or nozzle.

\subsection{Data sampling and processing} \label{s2c}

The anemometer signal was linearized, low-pass filtered, and then digitally sampled. We set the sampling frequency as high as possible, on the condition that high-frequency noise was not seen in the power spectrum. The filter cutoff was at one-half of the sampling frequency. We obtained a long record of $4 \times 10^8$ data in each experiment, except for $1 \times 10^8$ data in B1.

The anemometer signal is proportional to the flow velocity, through the calibration coefficient that depends on the condition of the anemometer and thereby varied slowly in time. For individual segments of each data record, the length of which is fixed for the record and ranges from $4 \times 10^6$ to $2 \times 10^7$ data, we determined the coefficient so as to have the same $U$ value. The coefficient within each segment is estimated to have varied by $\pm 1$\% at most.

We converted $u(t)$ and $v(t)$ at time $t$ into $u(x)$ and $v(x)$ at position $x$, by using Taylor's frozen-eddy hypothesis, $x = -Ut$, which requires that $\langle u^2 \rangle ^{1/2}/U$ is small enough. This requirement was satisfied in our experiments where $\langle u^2 \rangle ^{1/2}/U \lesssim 0.2$. Even at $\langle u^2 \rangle ^{1/2}/U \simeq 0.3$, Taylor's hypothesis holds valid as demonstrated by data obtained simultaneously with two probes separated by streamwise distances.\cite{sd98}

Since $u(t)$ and $v(t)$ are stationary, $u(x)$ and $v(x)$ are homogeneous, although actual turbulence in the wind tunnel was not homogeneous along the $x$ direction. Over small scales, fluctuations of $u(x)$ and $v(x)$ correspond to spatial fluctuations that actually existed in the wind tunnel. Those over large scales do not. They are interpreted as fluctuations over long timescales described in terms of large length scales.\cite{note2}

To calculate small-scale statistics such as the mean rate of energy dissipation $\langle \varepsilon \rangle$, we used $\partial_x v$ instead of usual $\partial_x u $ by assuming local isotropy, $\langle ( \partial_x v )^2 \rangle = 2 \langle ( \partial_x u )^2 \rangle$. Over smallest scales, the $u$ component measured by a crossed-wire probe is contaminated with the $w$ component that is perpendicular to the two wires of the probe.\cite{note3} The $v$ component is free from such contamination.

\begingroup
\squeezetable
\begin{table}
\caption{\label{t2} 
Statistics of $\ln \delta u_{r,R}^2$ and $\ln \delta v_{r,R}^2$ at $r/L_u = 0.01$, 0.03, and 0.1 as well as of $\ln \delta u_R^2$ and $\ln \delta v_R^2$ among segments with $R/L_u = 10$ for grid turbulence (G1 and G2), boundary layers (B1 and B2), and jets (J1 and J2). We also show the number of the segments used for the analyses.}

\begin{ruledtabular}
\begin{tabular}{lrrrrrr}
Quantity                     &  G1    &  G2    &  B1    &  B2    &  J1    &  J2    \\ 
\hline  
\noalign{\smallskip}
\multicolumn{7}{c}{Standard deviation}                                             \\
$\ln \delta u_{0.01Lu,R}^2$  &$ 0.15$ &$ 0.14$ &$ 0.17$ &$ 0.15$ &$ 0.11$ &$ 0.11$ \\
$\ln \delta u_{0.03L_u,R}^2$ &$ 0.14$ &$ 0.14$ &$ 0.16$ &$ 0.16$ &$ 0.12$ &$ 0.12$ \\
$\ln \delta u_{0.1L_u,R}^2$  &$ 0.16$ &$ 0.16$ &$ 0.18$ &$ 0.18$ &$ 0.14$ &$ 0.14$ \\
$\ln u_R^2$                  &$ 0.31$ &$ 0.32$ &$ 0.35$ &$ 0.35$ &$ 0.25$ &$ 0.26$ \\              
$\ln \delta v_{0.01L_u,R}^2$ &$ 0.15$ &$ 0.14$ &$ 0.16$ &$ 0.14$ &$ 0.12$ &$ 0.12$ \\
$\ln \delta v_{0.03L_u,R}^2$ &$ 0.14$ &$ 0.14$ &$ 0.15$ &$ 0.15$ &$ 0.13$ &$ 0.13$ \\
$\ln \delta v_{0.1L_u,R}^2$  &$ 0.16$ &$ 0.16$ &$ 0.17$ &$ 0.17$ &$ 0.15$ &$ 0.15$ \\
$\ln v_R^2$                  &$ 0.22$ &$ 0.23$ &$ 0.25$ &$ 0.25$ &$ 0.20$ &$ 0.20$ \\              
\noalign{\smallskip}
\multicolumn{7}{c}{Skewness factor}                                                \\
$\ln \delta u_{0.01L_u,R}^2$ &$-0.00$ &$+0.06$ &$-0.03$ &$+0.02$ &$-0.03$ &$-0.06$ \\
$\ln \delta u_{0.03L_u,R}^2$ &$+0.00$ &$+0.06$ &$-0.02$ &$+0.04$ &$-0.05$ &$-0.02$ \\
$\ln \delta u_{0.1L_u,R}^2$  &$-0.02$ &$+0.05$ &$-0.00$ &$-0.01$ &$-0.02$ &$-0.04$ \\
$\ln u_R^2$                  &$+0.19$ &$+0.22$ &$+0.19$ &$+0.08$ &$-0.02$ &$-0.00$ \\              
$\ln \delta v_{0.01L_u,R}^2$ &$-0.01$ &$+0.04$ &$-0.13$ &$-0.07$ &$-0.03$ &$-0.05$ \\
$\ln \delta v_{0.03L_u,R}^2$ &$+0.01$ &$+0.05$ &$-0.07$ &$-0.05$ &$-0.04$ &$-0.05$ \\
$\ln \delta v_{0.1L_u,R}^2$  &$+0.00$ &$+0.03$ &$-0.07$ &$-0.07$ &$-0.02$ &$-0.07$ \\
$\ln v_R^2$                  &$+0.03$ &$+0.05$ &$+0.01$ &$+0.01$ &$-0.06$ &$-0.01$ \\              
\noalign{\smallskip}
\multicolumn{7}{c}{Flatness factor}                                                \\
$\ln \delta u_{0.01L_u,R}^2$ &$ 2.94$ &$ 3.00$ &$ 3.02$ &$ 3.13$ &$ 3.02$ &$ 3.07$ \\
$\ln \delta u_{0.03L_u,R}^2$ &$ 2.95$ &$ 3.01$ &$ 3.00$ &$ 3.09$ &$ 3.06$ &$ 3.06$ \\
$\ln \delta u_{0.1L_u,R}^2$  &$ 2.97$ &$ 3.01$ &$ 3.02$ &$ 3.09$ &$ 3.02$ &$ 2.98$ \\
$\ln u_R^2$                  &$ 3.02$ &$ 3.03$ &$ 3.21$ &$ 2.95$ &$ 3.11$ &$ 2.95$ \\              
$\ln \delta v_{0.01L_u,R}^2$ &$ 2.95$ &$ 3.00$ &$ 3.11$ &$ 3.14$ &$ 3.04$ &$ 2.98$ \\
$\ln \delta v_{0.03L_u,R}^2$ &$ 2.97$ &$ 3.01$ &$ 3.07$ &$ 3.10$ &$ 3.00$ &$ 3.05$ \\
$\ln \delta v_{0.1L_u,R}^2$  &$ 2.99$ &$ 3.01$ &$ 3.03$ &$ 3.01$ &$ 2.97$ &$ 3.03$ \\
$\ln v_R^2$                  &$ 3.00$ &$ 3.03$ &$ 2.95$ &$ 2.97$ &$ 2.93$ &$ 2.91$ \\              
\noalign{\smallskip}
\multicolumn{7}{c}{Correlation coefficient against $\ln v_R^2$}                    \\
$\ln \delta u_{0.01L_u,R}^2$ &$+0.34$ &$+0.35$ &$+0.37$ &$+0.37$ &$+0.36$ &$+0.36$ \\
$\ln \delta u_{0.03L_u,R}^2$ &$+0.32$ &$+0.33$ &$+0.37$ &$+0.35$ &$+0.36$ &$+0.36$ \\
$\ln \delta u_{0.1L_u,R}^2$  &$+0.28$ &$+0.29$ &$+0.36$ &$+0.32$ &$+0.31$ &$+0.33$ \\
$\ln u_R^2$                  &$+0.10$ &$+0.12$ &$-0.02$ &$-0.05$ &$-0.04$ &$-0.03$ \\              
$\ln \delta v_{0.01L_u,R}^2$ &$+0.36$ &$+0.37$ &$+0.27$ &$+0.30$ &$+0.32$ &$+0.33$ \\
$\ln \delta v_{0.03L_u,R}^2$ &$+0.39$ &$+0.39$ &$+0.31$ &$+0.33$ &$+0.36$ &$+0.36$ \\
$\ln \delta v_{0.1L_u,R}^2$  &$+0.49$ &$+0.49$ &$+0.41$ &$+0.41$ &$+0.50$ &$+0.49$ \\
\noalign{\smallskip}
\multicolumn{7}{c}{Number of segments}                                             \\
$N_R$                        &$72600$ &$67600$ &$ 6360$ &$17560$ &$ 8120$ &$ 8000$ \\

\end{tabular}
\end{ruledtabular}
\end{table}
\endgroup
\begin{figure}[b]
\resizebox{7.8cm}{!}{\includegraphics*[5cm,4.25cm][16cm,27.2cm]{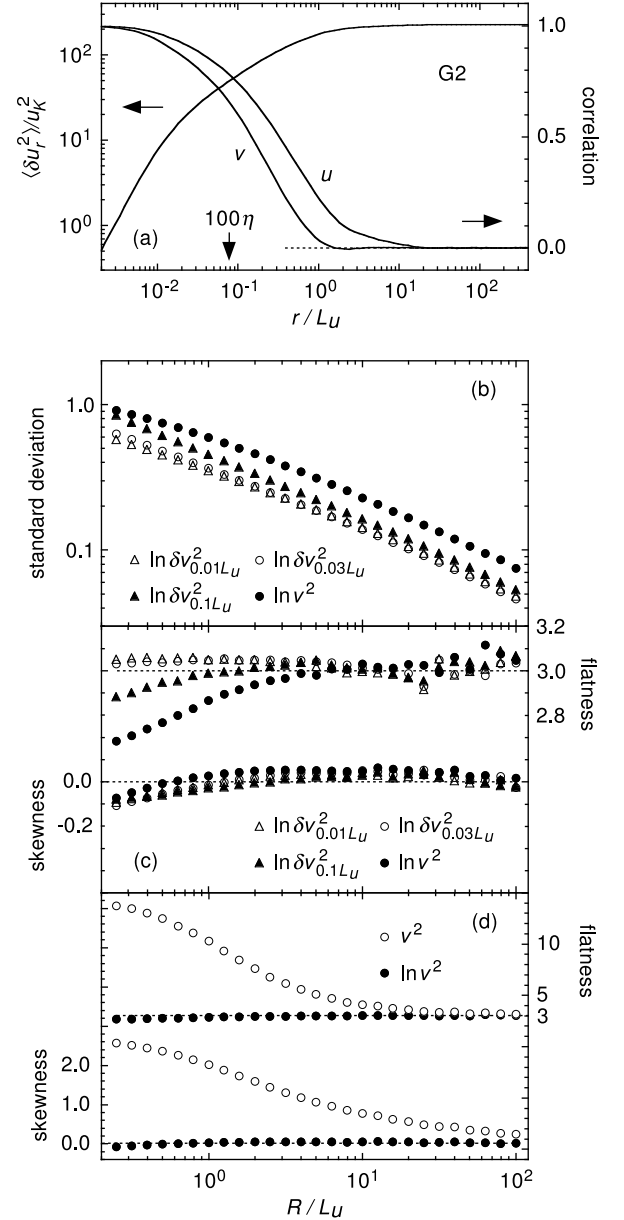}}
\caption{\label{f1} Statistics for grid turbulence G2 as a function of $r/L_u$ or $R/L_u$. (a) $\langle u(x+r)u(x) \rangle /\langle u^2 \rangle$, $\langle v(x+r)v(x) \rangle /\langle v^2 \rangle$, and $\langle \delta u_r^2 \rangle/u_K^2$. The arrow indicates the scale $r = 100 \eta$. (b) Standard deviations of $\ln \delta v_{0.01L_u,R}^2$ ($\vartriangle$), $\ln \delta v_{0.03L_u,R}^2$ ($\circ$), $\ln \delta v_{0.1L_u,R}^2$ ($\blacktriangle$), and $\ln v_R^2$ ($\bullet$). (c) Skewness and flatness factors of $\ln \delta v_{0.01L_u,R}^2$ ($\vartriangle$), $\ln \delta v_{0.03L_u,R}^2$ ($\circ$), $\ln \delta v_{0.1L_u,R}^2$ ($\blacktriangle$), and $\ln v_R^2$ ($\bullet$). (d) Skewness and flatness factors of $v_R^2$ ($\circ$) and $\ln v_R^2$ ($\bullet$).}
\end{figure} 
\begin{figure}[b]
\resizebox{7.8cm}{!}{\includegraphics*[5cm,4.25cm][16cm,27.2cm]{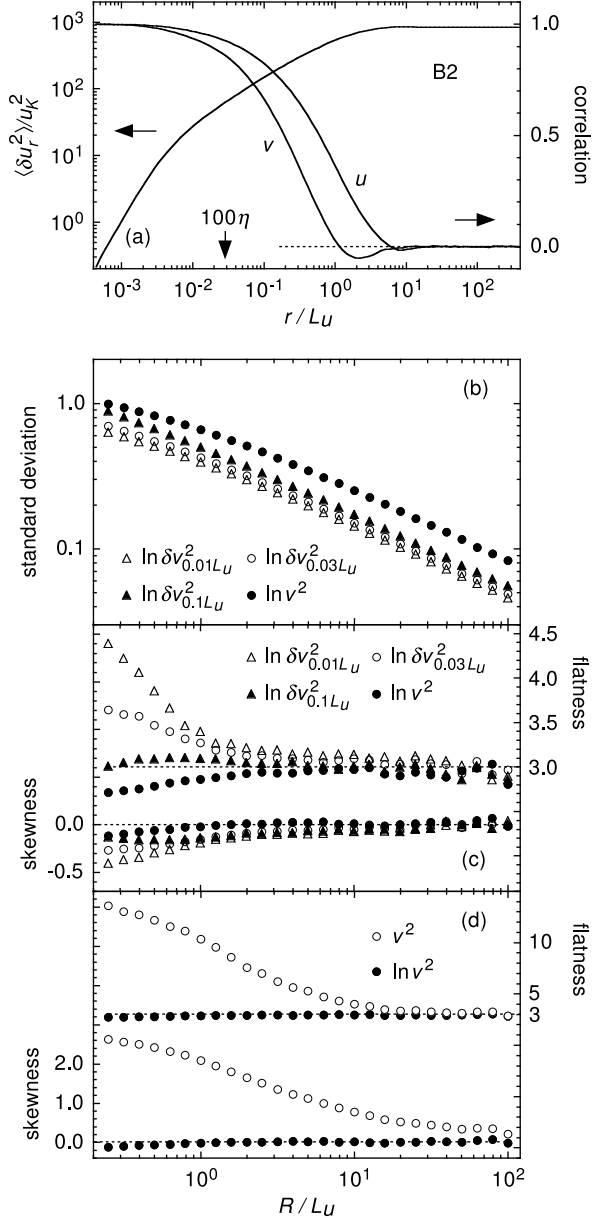}}
\caption{\label{f2} Same as in Fig. \ref{f1} but for boundary layer B2.}
\end{figure} 
\begin{figure}[b]
\resizebox{7.8cm}{!}{\includegraphics*[5cm,4.25cm][16cm,27.2cm]{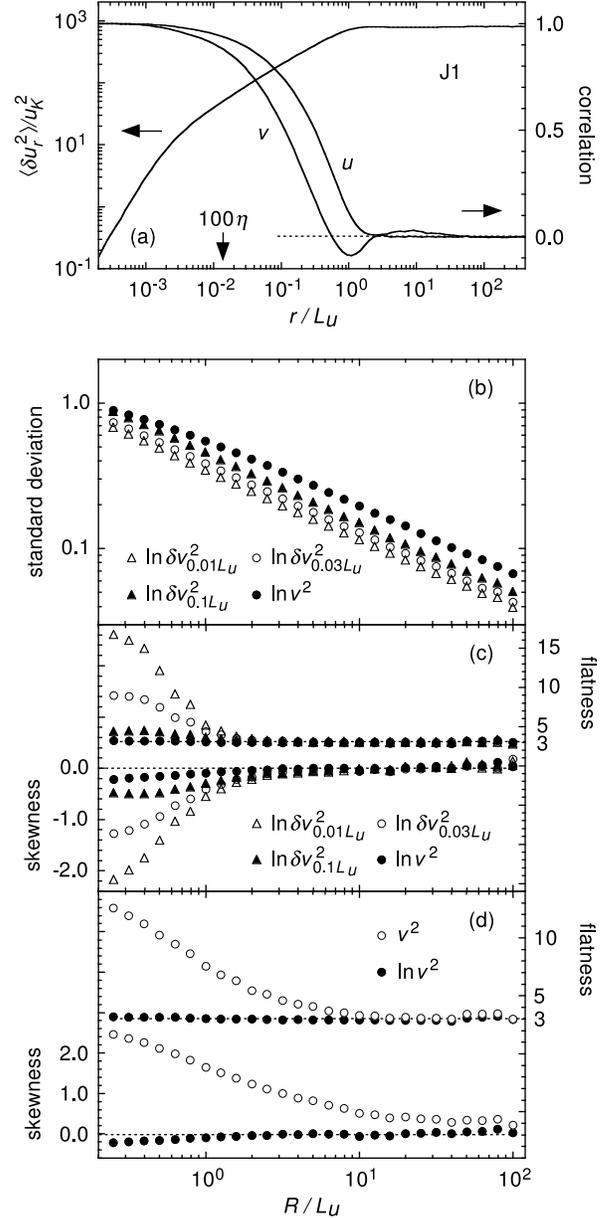}}
\caption{\label{f3} Same as in Fig. \ref{f1} but for turbulent jet J1.}
\end{figure} 

\section{RESULTS AND DISCUSSION} \label{s3}

The data records are divided into segments with length $R$. For each segment, the center of which is defined as $x_c$, we calculate the mean energies at scale $r < R$ as
\begin{subequations}
\begin{align}
&\delta u_{r,R}^2(x_c) = \frac{1}{R-r} \int ^{x_c+R/2-r}_{x_c-R/2} \delta u_r^2(x) dx,\\
&\delta v_{r,R}^2(x_c) = \frac{1}{R-r} \int ^{x_c+R/2-r}_{x_c-R/2} \delta v_r^2(x) dx,
\end{align}
\end{subequations}with $\delta u_r(x) = u(x+r)-u(x)$ and $\delta v_r(x) = v(x+r)-v(x)$. The mean total energies are also calculated as
\begin{subequations}
\begin{align}
&u^2_R(x_c)         = \frac{1}{R}   \int ^{x_c+R/2}  _{x_c-R/2} u^2 (x) dx,        \\
&v^2_R(x_c)         = \frac{1}{R}   \int ^{x_c+R/2}  _{x_c-R/2} v^2 (x) dx.        
\end{align}
\end{subequations} These are coarse-grained quantities where fluctuations at scales smaller than $R$ have been smoothed away. We study their statistics as a function of $R$. The results are summarized in Figs. \ref{f1}--\ref{f3} and Table \ref{t2}.

The segment length $R$ is given in units of the correlation length $L_u$. This is the representative of large scales. Up to about $10^0 L_u$, the correlations $\langle u(x+R)u(x) \rangle$ and $\langle v(x+R)v(x) \rangle$ are significant [Figs. \ref{f1}(a), \ref{f2}(a), and \ref{f3}(a)]. They are weak but not yet absent at about $10^0 L_u$--$10^2 L_u$, which is the $R$ range of our main interest. The scale $r$ is also given in units of $L_u$ as $0.01 L_u$, $0.03 L_u$ or $0.1 L_u$. At $r/L_u \rightarrow \infty$, $\delta u_{r,R}^2$ and $\delta v_{r,R}^2$ correspond to $u^2_R$ and $v^2_R$ in a statistical sense.

\subsection{Significance of large-scale fluctuations} \label{s3a}

Figures \ref{f1}(b), \ref{f2}(b), and \ref{f3}(b) show the standard deviations of $\ln \delta v_{r,R}^2$ and $\ln v^2_R$ as a function of $R/L_u$. For example, the standard deviation of $\ln v_R^2$ is
\begin{equation}
\langle (\ln v^2_R - \langle \ln v^2_R \rangle)^2 \rangle^{1/2}.
\end{equation}
The average is over all the segments. With an increase of $R$, the standard deviations decrease, but they are not so small even at largest $R$. Thus, large-scale fluctuations of $\delta v_r^2$ and $v^2$ are significant.\cite{m06,m08} The same is true for $\delta u_r^2$ and $u^2$ (Table \ref{t2}).

If we fix $R/L_u$ and $r/L_u$, the standard deviations of $\ln \delta u_{r,R}^2$, $\ln \delta v_{r,R}^2$, and $\ln v^2_R$ are almost the same among the datasets. This implies the existence of some common process. The exception is $\ln u^2_R$ that suffers from largest scale motions of the flow (Sec. \ref{s2b}). With an increase of $r$, the standard deviations of $\ln \delta u_{r,R}^2$ and $\ln \delta v_{r,R}^2$ tend to increase.

\subsection{Lognormality of large-scale fluctuations} \label{s3b}

Figures \ref{f1}(c), \ref{f2}(c), and \ref{f3}(c) show the skewness and flatness factors of $\ln \delta v_{r,R}^2$ and $\ln v^2_R$. For example, the skewness factor of $\ln v^2_R$ is
\begin{subequations}
\begin{equation}
\frac{ \langle (\ln v^2_R - \langle \ln v^2_R \rangle)^3 \rangle}
     { \langle (\ln v^2_R - \langle \ln v^2_R \rangle)^2 \rangle^{3/2}}.
\end{equation}
The flatness factor of $\ln v^2_R$ is
\begin{equation}
\frac{ \langle (\ln v^2_R - \langle \ln v^2_R \rangle)^4 \rangle}
     { \langle (\ln v^2_R - \langle \ln v^2_R \rangle)^2 \rangle^2}.
\end{equation}
\end{subequations} At $R/L_u \simeq 10^0$--$10^2$, the skewness and flatness factors are close to the Gaussian values of 0 and 3. Thus, large-scale fluctuations of $\delta v_r^2$ and $v^2$ are lognormal. The lognormality is expected to be universal because it is observed in all the datasets. They are different in the Reynolds number or the flow configuration. Also for $\delta u_{r,R}^2$ (Table \ref{t2}), we observe the lognormality.

\begin{figure}[b]
\resizebox{7.8cm}{!}{\includegraphics*[5cm,7.75cm][16cm,27cm]{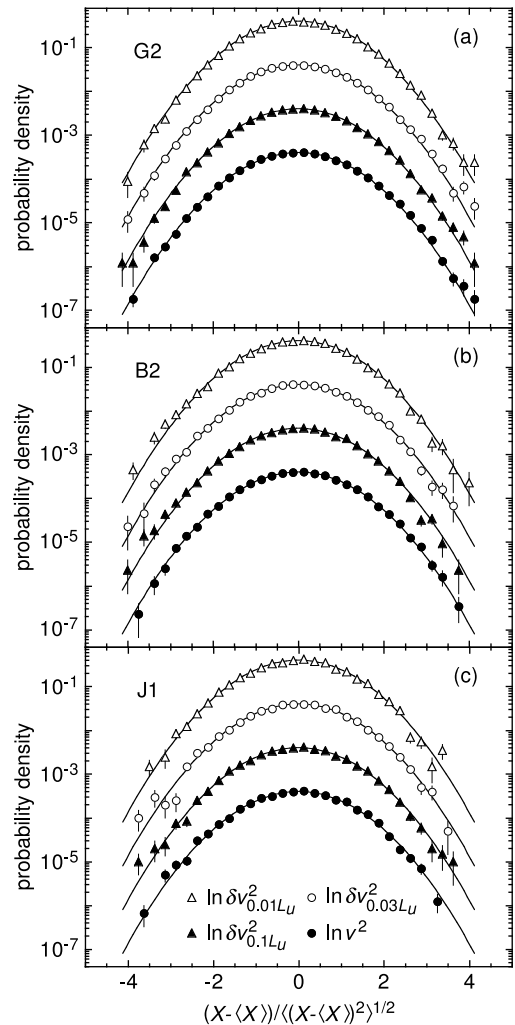}}
\caption{\label{f4} Probability densities $P(X_R)$ of $X_R = \ln \delta v_{0.01L_u,R}^2$ ($\vartriangle$), $\ln \delta v_{0.03L_u,R}^2$ ($\circ$), $\ln \delta v_{0.1L_u,R}^2$ ($\blacktriangle$), and $\ln v_R^2$ ($\bullet$) at $R/L_u = 10$ as a function of $(X_R - \langle X_R \rangle)/\langle (X_R - \langle X_R \rangle)^2 \rangle^{1/2}$. The error bars represent $P(X_R)/N_{X_R}^{1/2}$, where $N_{X_R}$ is the number of segments that fall in the bin. The curves represent Gaussian distributions. They are individually shifted by a factor $10$. (a) Grid turbulence G2. (b) Boundary layer B2. (c) Turbulent jet J1.}
\end{figure}

Figure \ref{f4} shows the probability density distributions of $\ln \delta v_{r,R}^2$ and $\ln v^2_R$ at $R/L_u = 10$. The distributions are Gaussian, at least in the observed range that extends to several standard deviations from the peak.

In some datasets, deviation from the lognormality is seen for $u_R^2$ (Table \ref{t2}). It tends toward the lognormality but suffers from largest scale motions of the flow (Sec. \ref{s2b}).

The lognormal fluctuations observed here are distinct from those modelled by Kolmogorov.\cite{k62} While he was interested in small-scale intermittency\cite{sa97} and studied $\delta u_r^n$ at small $r$ to obtain their scaling laws, we are interested in large-scale fluctuations and study $\delta u_{r,R}^n$ at large $R$. The scaling laws of $\delta u_{r,R}^n$ are not necessary. Thus, our study is not necessarily concerned with the well-known problems of Kolmogorov's lognormal model, e.g., violation of Novikov's inequality\cite{n71} for scaling exponents. In addition, we are not intended to argue that the lognormality is exact (Sec. \ref{s3e}).

\subsection{Correlation among large-scale fluctuations} \label{s3c}

Table \ref{t2} shows the correlation coefficients of $\ln \delta u_{r,R}^2$, $\ln \delta v_{r,R}^2$, and $\ln u^2_R$ against $\ln v_R^2$. For example, the coefficient of $\ln \delta v_{r,R}^2$ is
\begin{equation}
\frac{ \langle (\ln \delta v^2_{r,R} - \langle \ln \delta v^2_{r,R} \rangle)(\ln v^2_R - \langle \ln v^2_R \rangle) \rangle }
     { \langle (\ln \delta v^2_{r,R} - \langle \ln \delta v^2_{r,R} \rangle)^2 \rangle^{1/2}
     \langle (\ln v^2_R - \langle \ln v^2_R \rangle)^2 \rangle^{1/2} }.
\end{equation}
The correlations are significant for most of the quantities. They tend to fluctuate synchronously.

The datasets share interesting features. Between $\ln u_R^2$ and $\ln v_R^2$, the correlation is weak or even absent. With a decrease of $r$, the correlation of $\ln \delta u_{r,R}^2$ against $\ln v_R^2$ is increasingly strong, while that of $\ln \delta v_{r,R}^2$ is increasingly weak. Taking also account of the observed decrease of the standard deviations with a decrease of $r$ (Sec. \ref{s3a}), we consider that the fluctuations originate in large $r$. Those in the $u$ and $v$ components are originally uncorrelated but become correlated in the course of their propagation to the smaller $r$.

\begingroup
\squeezetable
\begin{table}
\caption{\label{t3} 
Same as in Table \ref{t2} but for $\ln \delta u_{r,R}^2$ and $\ln \delta v_{r,R}^2$ at $r/\eta = 30$, 100, and 300 as well as for $\ln \varepsilon_R$.}

\begin{ruledtabular}
\begin{tabular}{lrrrrrr}
Quantity                      &  G1    &  G2    &  B1    &  B2    &  J1    &  J2    \\ 
\hline  
\noalign{\smallskip}
\multicolumn{7}{c}{Standard deviation}                                              \\
$\ln \delta u_{30\eta,R}^2$   &$ 0.14$ &$ 0.14$ &$ 0.16$ &$ 0.16$ &$ 0.12$ &$ 0.11$ \\
$\ln \delta u_{100\eta,R}^2$  &$ 0.16$ &$ 0.15$ &$ 0.17$ &$ 0.16$ &$ 0.12$ &$ 0.12$ \\
$\ln \delta u_{300\eta,R}^2$  &$ 0.20$ &$ 0.19$ &$ 0.19$ &$ 0.17$ &$ 0.13$ &$ 0.13$ \\
$\ln \varepsilon_R$           &$ 0.16$ &$ 0.16$ &$ 0.20$ &$ 0.19$ &$ 0.14$ &$ 0.13$ \\
$\ln \delta v_{30\eta,R}^2$   &$ 0.14$ &$ 0.14$ &$ 0.15$ &$ 0.14$ &$ 0.11$ &$ 0.11$ \\
$\ln \delta v_{100\eta,R}^2$  &$ 0.16$ &$ 0.16$ &$ 0.16$ &$ 0.15$ &$ 0.12$ &$ 0.12$ \\
$\ln \delta v_{300\eta,R}^2$  &$ 0.20$ &$ 0.19$ &$ 0.19$ &$ 0.17$ &$ 0.13$ &$ 0.13$ \\
\noalign{\smallskip}
\multicolumn{7}{c}{Skewness factor}                                                 \\
$\ln \delta u_{30\eta,R}^2$   &$+0.00$ &$+0.06$ &$-0.02$ &$+0.00$ &$-0.03$ &$-0.07$ \\
$\ln \delta u_{100\eta,R}^2$  &$-0.02$ &$+0.06$ &$-0.01$ &$+0.04$ &$-0.03$ &$-0.05$ \\
$\ln \delta u_{300\eta,R}^2$  &$-0.02$ &$+0.01$ &$-0.01$ &$+0.00$ &$-0.04$ &$-0.01$ \\
$\ln \varepsilon_R$           &$-0.00$ &$+0.05$ &$-0.15$ &$-0.23$ &$-0.04$ &$-0.10$ \\
$\ln \delta v_{30\eta,R}^2$   &$+0.01$ &$+0.05$ &$-0.09$ &$-0.09$ &$-0.04$ &$-0.08$ \\
$\ln \delta v_{100\eta,R}^2$  &$-0.00$ &$+0.04$ &$-0.06$ &$-0.05$ &$-0.03$ &$-0.05$ \\
$\ln \delta v_{300\eta,R}^2$  &$-0.02$ &$+0.01$ &$-0.05$ &$-0.07$ &$-0.03$ &$-0.06$ \\
\noalign{\smallskip}
\multicolumn{7}{c}{Flatness factor}                                                 \\
$\ln \delta u_{30\eta,R}^2$   &$ 2.95$ &$ 3.00$ &$ 3.03$ &$ 3.13$ &$ 3.00$ &$ 3.05$ \\
$\ln \delta u_{100\eta,R}^2$  &$ 2.97$ &$ 3.01$ &$ 3.04$ &$ 3.09$ &$ 3.02$ &$ 3.06$ \\
$\ln \delta u_{300\eta,R}^2$  &$ 2.98$ &$ 2.99$ &$ 3.01$ &$ 3.07$ &$ 3.05$ &$ 3.05$ \\
$\ln \varepsilon_R$           &$ 2.94$ &$ 2.98$ &$ 3.17$ &$ 3.35$ &$ 3.04$ &$ 2.99$ \\
$\ln \delta v_{30\eta,R}^2$   &$ 2.97$ &$ 3.01$ &$ 3.06$ &$ 3.15$ &$ 3.02$ &$ 2.97$ \\
$\ln \delta v_{100\eta,R}^2$  &$ 2.99$ &$ 3.01$ &$ 3.03$ &$ 3.10$ &$ 3.04$ &$ 2.99$ \\
$\ln \delta v_{300\eta,R}^2$  &$ 2.97$ &$ 3.01$ &$ 3.06$ &$ 3.01$ &$ 2.97$ &$ 3.05$ \\
\noalign{\smallskip}
\multicolumn{7}{c}{Correlation coefficient against $\ln v_R^2$}                     \\
$\ln \delta u_{30\eta,R}^2$   &$+0.31$ &$+0.33$ &$+0.37$ &$+0.37$ &$+0.36$ &$+0.36$ \\
$\ln \delta u_{100\eta,R}^2$  &$+0.28$ &$+0.30$ &$+0.36$ &$+0.35$ &$+0.36$ &$+0.36$ \\
$\ln \delta u_{300\eta,R}^2$  &$+0.23$ &$+0.25$ &$+0.32$ &$+0.33$ &$+0.35$ &$+0.35$ \\
$\ln \varepsilon_R$           &$+0.35$ &$+0.36$ &$+0.23$ &$+0.25$ &$+0.29$ &$+0.29$ \\
$\ln \delta v_{30\eta,R}^2$   &$+0.39$ &$+0.38$ &$+0.29$ &$+0.30$ &$+0.32$ &$+0.32$ \\
$\ln \delta v_{100\eta,R}^2$  &$+0.49$ &$+0.46$ &$+0.36$ &$+0.33$ &$+0.33$ &$+0.33$ \\
$\ln \delta v_{300\eta,R}^2$  &$+0.65$ &$+0.60$ &$+0.52$ &$+0.40$ &$+0.38$ &$+0.37$ \\
\noalign{\smallskip}
\multicolumn{7}{c}{Number of segments}                                              \\
$N_R$                         &$72600$ &$67600$ &$ 6360$ &$17560$ &$ 8120$ &$ 8000$ \\

\end{tabular}
\end{ruledtabular}
\end{table}
\endgroup

\subsection{Large-scale fluctuations of small-scale quantities} \label{s3d}

We have studied $\delta u_{r,R}^2$ and $\delta v_{r,R}^2$ at $r/L_u = 0.01$, 0.03, and 0.1. They are now studied at $r/\eta = 30$, 100, and 300. The Kolmogorov length $\eta$ is the representative of small scales. Judging from $\langle \delta u_r^2 \rangle$ in Figs. \ref{f1}(a), \ref{f2}(a), and \ref{f3}(a), the scale $30\eta$ lies in the dissipation subrange. The scale $100\eta$ lies in the inertial subrange with $\langle \delta u_r^2 \rangle \propto r^{2/3}$. This subrange extends beyond $300 \eta$ if the Reynolds number is high enough. We also study the energy dissipation rate averaged over each segment
\begin{equation}
\varepsilon_R(x_c) = \frac{1}{R}   \int ^{x_c+R/2}  _{x_c-R/2} \varepsilon (x) dx,
\end{equation}
with $\varepsilon = 15\nu (\partial_x v)^2/2$.\cite{note4} It corresponds to $\delta v_r^2$ at $r/\eta \rightarrow 0$.

Table \ref{t3} shows the results at $R/L_u = 10$. The fluctuations are lognormal. The correlations against $\ln v_R^2$ have the same features as in Table \ref{t2}. However, compared with the standard deviations at fixed $r/L_u$ in Table \ref{t2}, those at fixed $r/\eta$ in Table \ref{t3} tend to scatter among the datasets. Therefore, the fluctuations are described in terms of the correlation length $L_u$ that represents large scales, rather than in terms of the Kolmogorov length $\eta$ that represents small scales (see also Sec. \ref{s3c}).

The standard deviation of $\ln \varepsilon_R$ is larger than that of $\ln \delta v_{30\eta,R}^2$, although the standard deviation of $\ln \delta v_{r,R}^2$ decreases with a decrease of $r$ down to $30\eta$. Probably, $\ln \varepsilon_R$ suffers from small-scale intermittency that causes a broad $\varepsilon$ distribution.\cite{sa97}

For $\varepsilon_R$ in the dataset B1 and B2, we see deviation from the lognormality. This is achieved at the larger $R$. In fact, at $R/L_u = 100$, the skewness factor is $+0.03$ (B1) or $+0.00$ (B2). The flatness factor is 3.04 (B1) or 3.00 (B2).

\subsection{Origin of large-scale lognormal fluctuations} \label{s3e}

Regardless of the Reynolds number and the flow configuration, the energies $\delta u_r^2$ and $\delta v_r^2$ at scale $r$, the total energies $u^2$ and $v^2$, and also the energy dissipation rate $\varepsilon$ fluctuate over large scales in lognormal distributions. Their standard deviations are almost the same among the flows (Secs. \ref{s3a}, \ref{s3b}, and \ref{s3d}). Hence, some common process is at work, which is discussed here.

The observed fluctuations originate in turbulence itself. They are too strong to be an experimental artifact, i.e., some variation in the condition of the wind tunnel or the measurement system. In addition, if such a variation were responsible for the fluctuations, they should exhibit a correlation between $u_R^2$ and $v_R^2$ that is actually not significant at all (Sec. \ref{s3c}).

A lognormal fluctuation is induced by some multiplicative stochastic process, e.g., a product of many independent stochastic variables, $Q_N = \prod_{n=1}^N q_n$. Its logarithm is a sum of many independent stochastic variables, $\ln Q_N = \sum_{n=1}^N \ln q_n$, which is Gaussian according to the central limit theorem. For the lognormal fluctuations observed here, the process is related to interactions among scales through the energy transfer.\cite{m08} While the mean energy transfer is to a smaller scale and is significant between scales in the inertial subrange alone, the local energy transfer is either to a smaller or larger scale and is significant between all scales.\cite{m06,dr90,ok92,mininni06} Thus, any scale interacts with itself and with many other scales.

The fluctuations observed for different scales $r$ tend to be synchronous (Sec. \ref{s3c}). This is consistent with the above explanation that the fluctuations are induced by interactions among scales.

The rate of the energy transfer does not have a lognormal fluctuation because the rate could change its sign. Although the fluctuation of this rate is important, it is beyond the reach of our experimental study.

To apply the central limit theorem, there exist various forms of sufficient but not necessary conditions.\cite{f71} The essence is that the stochastic variables are all alike and no few of them dominate the others. It is also required that, among the variables, correlations are absent exactly or at least practically. These requirements have to be satisfied by the interactions among scales that are to reproduce the observed lognormality.

The interactions are also required to involve a wide range of scales because the lognormal fluctuations are observed up to large $R$ and down to small $r$. Large scales are especially important as implied by the observed features of the standard deviations and correlations (Secs. \ref{s3a} and \ref{s3c}).

The central limit theorem does not determine the distribution at its tail, which might deviate from the lognormal distribution and even might depend on the Reynolds number or the flow configuration. Whether such a tail is prominent depends on the case. An extreme case could be the observed deviation from the lognormality of $u_R^2$ in some datasets (Sec. \ref{s3b}). Nevertheless, the lognormality is important at least as an idealized model because it exists only if the logarithm is subject to the central limit theorem. The lognormality also accounts for most of the distributions in the observed range (Fig. \ref{f4}). To observe the distributions over the wider range, much longer data are necessary.

\subsection{Scale range of lognormal fluctuations} \label{s3f}

The lognormal fluctuations exist at $R/L_u \simeq 10^0$--$10^2$, where the correlations $\langle u(x+R)u(x) \rangle$ and $\langle v(x+R)v(x) \rangle$ are weak but not yet absent [Figs. \ref{f1}(a), \ref{f2}(a), and \ref{f3}(a)]. Although the correlations there exhibit different profiles  among the flow configurations, this difference is not relevant to the existence of the lognormality. Outside the $R$ range, the correlations are either too strong or absent. The lognormal fluctuations do not exist.

At $R/L_u \lesssim 10^0$, the skewness and flatness factors of $\ln \delta v_{r,R}^2$ and $\ln v_R^2$ deviate from the Gaussian values [Figs. \ref{f1}(c), \ref{f2}(c), and \ref{f3}(c)]. The fluctuations are not lognormal and no longer explained by a multiple of independent processes. This is because the correlations $\langle u(x+R)u(x) \rangle$ and $\langle v(x+R)v(x) \rangle$ are too strong. The observed deviation from the lognormality is different among the dataset. Note that, among the figures, different are the ordinates for the skewness and flatness factors. Since this $R$ range for small scales has been studied in detail,\cite{sa97} we do not study it any more.

At $R/L_u \gtrsim 10^2$, the correlations $\langle u(x+R)u(x) \rangle$ and $\langle v(x+R)v(x) \rangle$ are absent. Then, any segment with length $R$ is divisible into many subsegments that are independent of one another. The length $R_*$ of these subsegments is such that $L_u \ll R_* \ll R$. By using a sum over the subsegments, we write $\varepsilon_R$, $\delta u_{r,R}^2$, $\delta v_{r,R}^2$, $u_R^2$, and $v_R^2$ as, e.g.,
\begin{equation} \label{eq7}
v_R^2(x_c) = \frac{R_*}{R} \sum_{n=1}^{R/R_*} v_{R_*}^2(x_{c_n}) .
\end{equation}
Here $x_{c_n}$ is the center of the $n$-th subsegment. The sum is subject to the central limit theorem and thereby fluctuates in a Gaussian distribution.\cite{m06,kg03} Figures \ref{f1}(d), \ref{f2}(d), and \ref{f3}(d) show the skewness and flatness factors of $v_R^2$, which in fact approach to the Gaussian values with an increase of $R/L_u$.\cite{note5,note6} In other words, when $R/L_u$ is large enough, quantities such as $R v_R^2$ are additive.\cite{m06} The value of an additive quantity for a region is the sum of its values for the subregions that are independent of one another.\cite{ll79} Such additive quantities are well known in thermodynamics and statistical mechanics, for which correlations are absent over scales of interest.

\section{CONCLUDING REMARKS} \label{s4}

For several flows of laboratory turbulence, we obtained long records of velocity data. These records were divided into numerous segments with length $R$. In each segment, we calculated the mean rate of energy dissipation $\varepsilon_R$, the mean energies $\delta u_{r,R}^2$ and $\delta v_{r,R}^2$ at scale $r$, and the mean total energies $u_R^2$ and $v_R^2$. Their values fluctuate significantly and synchronously. The fluctuations are lognormal. Since the lognormality is observed regardless of the Reynolds number and the configuration for turbulence production, it is expected to be universal.

The lognormal fluctuations are of large scales, $R/L_u \simeq 10^0$--$10^2$, where the correlations $\langle u(x+R)u(x) \rangle$ and $\langle v(x+R)v(x) \rangle$ are weak but not yet absent. This $R$ range had not been studied, although it lies in between the ranges that had been studied in detail.

We explained the lognormal fluctuations by applying the central limit theorem to the logarithm of a multiplicative stochastic process, which was related to interactions among many scales through the energy transfer. To apply the central limit theorem, it is required that the interactions are  all alike and do not have mutual correlations. In addition, features of the observed fluctuations require that the interactions involve a wide range of scales. Large scales are especially important.

The first discussion on large-scale fluctuations was by Landau,\cite{ll59} who argued against the universality of small-scale statistics discussed by Kolmogorov.\cite{k41} Kolmogorov discussed that small-scale statistics depend on $\nu$ and $\langle \varepsilon \rangle$ alone and are thus universal. Landau discussed that small-scale statistics are not universal because the large-scale $\varepsilon$ fluctuation is significant and is not universal. In practice, large-scale fluctuations are universally lognormal. Their standard deviations are also almost universal. Even if large scales affect small scales through the $\varepsilon$ fluctuation,\cite{o62,bc03} it does not necessarily violate the universality of small-scale statistics. However, past experiments suggested that some small-scale statistics are not universal and depend on the large-scale flow.\cite{m06,sd98,k92,p93,ss96} Since the Reynolds numbers of these studies might not have been high enough to distinguish between small and large scales,\cite{ks00,ab06} it is desirable to study the effect of large scales on small scales at the higher Reynolds numbers.

Finally, we note that the large-scale lognormal fluctuations are not restricted to those of turbulence. They should also exist in other nonlinear systems that have many degrees of freedom. This is because, regardless of the process details, lognormality exists only if the logarithm is subject to the central limit theorem. However, so far as we could see, no example is known. It is desirable to explore the lognormal fluctuations in various systems, promisingly over length scales or timescales where correlations are weak but not yet absent. We expect that the large-scale lognormal fluctuations turn out to be ubiquitous, far beyond those of turbulence observed here.

\acknowledgments

The authors are grateful to Y. Kawashima and K. Hashimoto for their continued collaboration.



\begin{thebibliography}{999}
\bibitem{note1} Compared with usual jet experiments, our measurement position was close to the nozzle exit with respect to its size and was also close to the floor of the test section. This fact is nevertheless not serious to our study, for which it is sufficient to prepare a variety of turbulent flows.
\bibitem{note2} The alternative interpretation is that such fluctuations correspond to spatial fluctuations of some virtual turbulence. Local values of $u(x)$ and $v(x)$ represent local regions of actual turbulence and are continuously connected up to largest scales (see also Appendix of Ref. \onlinecite{m06}).
\bibitem{note3} The two wires individually respond to all the $u$, $v$, and $w$ components. Since the measured $u$ component corresponds to the sum of the responses of the two wires, it is contaminated with the $w$ component. Since the measured $v$ component corresponds to the difference of the responses, it is not contaminated with the $w$ component.
\bibitem{note4} The velocity derivative $\partial _x v$ was estimated as $[ 8v(x+\delta x)-8v(x-\delta x)-v(x+2\delta x)+v(x-2\delta x) ] / 12\delta x$ with $\delta x=U/f_s$. Even if it is estimated as $[v(x+\delta x)-v(x)] / \delta x$, the results are essentially the same.
\bibitem{note5} Even at $R/L_u \gtrsim 10^2$, the skewness and flatness factors of $\ln v_R^2$ remain close to the Gaussian values. The reason is that, if the standard deviation is small enough, a Gaussian distribution is approximated by a lognormal distribution. We underline that the lognormality observed at $R/L_u \lesssim 10^2$ for $v_R^2$ is not due to such an approximation. This has been ascertained also for $\delta v_{r,R}^2$ and so on.
\bibitem{note6} The other implication of Eq. (\ref{eq7}) is that the standard deviation of $v_R^2$ scales as $R^{-1/2}$ (Ref. \onlinecite{m06}). If this standard deviation is small enough, the Gaussian distribution of $v_R^2$ is approximated by a lognormal distribution (Note \onlinecite{note5}), and hence the standard deviation of $\ln v_R^2$ also scales as, albeit approximately, $R^{-1/2}$. We see such scalings at $R/L_u \gtrsim 10^2$ in Figs. \ref{f1}(b), \ref{f2}(b), and \ref{f3}(b).
\end{thebibliography}
\end{document}